\documentclass[aps,prd,10pt,a4paper,english,amssymb,nofootinbib,superscriptaddress,twocolumn]{revtex4-2}

\usepackage[utf8]{inputenc}
\usepackage[T1]{fontenc}
\usepackage{lmodern}

\usepackage{amsmath,amssymb,bm,esint}
\usepackage{latexsym}
\usepackage{xcolor}
\definecolor{RED}{rgb}{1,0,0}

\usepackage{graphicx}
\usepackage{subfigure}

\usepackage[unicode=true,pdfusetitle,
  bookmarks=true,bookmarksnumbered=false,bookmarksopen=false,
  breaklinks=false,pdfborder={0 0 1},backref=false,colorlinks=true]{hyperref}
\hypersetup{linkcolor=blue,citecolor=blue}

\begin{document}

\title{Geometrically Significant Surfaces of Black Holes from a Single Scalar}

\author{Cagdas Ulus Agca}
\email{ulusagca@metu.edu.tr}
\affiliation{{\small Department of Physics,}\\ {\small Middle East Technical University, 06800 Ankara, Turkey}}

\author{Bayram Tekin}
\email{bayram.tekin@bilkent.edu.tr}
\affiliation{{\small Department of Physics,}\\ {\small Bilkent University, 06800 Ankara, Turkey}}

\begin{abstract}
Black hole spacetimes contain several geometrically distinguished hypersurfaces, including event and Cauchy horizons, stationary-limit surfaces, and curvature singularities. These structures are usually identified by different geometric or causal criteria. We show that, for the Kerr--Newman black hole, the membrane-paradigm pressure of a stretched horizon admits an analytically continued scalar representative whose fully factorized form has zeros, poles, and singular factors aligned with these standard Kerr--Newman loci. Its zeros occur at the outer and inner horizons, its poles occur at the outer and inner stationary-limit surfaces, its higher-order divergence contains the ring-singularity factor, and its large-$r$ behavior records the asymptotic decay of the scalar. The construction is not proposed as a new invariant characterization of the spacetime, nor as a physical extension of the membrane fluid into the black-hole interior. Rather, it gives a compact membrane-pressure-based diagnostic whose analytic structure reorganizes several familiar surfaces of the Kerr--Newman geometry. We also note a secondary, purely algebraic analogy with generalized multi-component van der Waals-type equations of state.
\end{abstract}

\maketitle
\section{Introduction}
Black hole spacetimes contain several geometrically and physically distinguished hypersurfaces: event horizons, Cauchy horizons, stationary-limit (ergo) surfaces, curvature singularities, and asymptotic infinity. In the standard treatment, these structures are identified by different criteria: horizons by null generators or global causal constructions; ergosurfaces by the vanishing of the stationary Killing norm; singularities by geodesic incompleteness or by the divergence of curvature invariants; and asymptotic flatness by the large-distance behavior of the metric. Thus, even within a single exact solution, the critical surfaces of the spacetime are usually detected by different geometric probes. Historically, the significance of these surfaces was also understood in stages: the invariant meaning of the Schwarzschild horizon was clarified much later than the original solution \cite{Finkelstein}, the physical role of the ergoregion emerged with the Kerr geometry \cite{Kerr,Penrose}, the inner horizon was identified as a Cauchy horizon by Carter \cite{Carter}, and Penrose emphasized that the true singularity is fundamentally tied to geodesic incompleteness rather than merely to the divergence of invariants \cite{Penrose2}.

In this article, we show that for the Kerr--Newman black hole there is a simple scalar representative, obtained from the membrane-paradigm pressure by analytic continuation, whose factorized analytic structure is aligned with these distinguished loci. More precisely, its zeros identify the outer and inner horizons, its poles identify the outer and inner stationary-limit surfaces, and its singular denominator contains the ring-singularity factor. The point is not that these surfaces were previously unknown, nor that the scalar gives a new invariant classification of Kerr--Newman geometry. The point is that a quantity naturally inherited from the stretched-horizon membrane stress tensor reorganizes several standard pieces of Kerr--Newman geometry into one compact expression.

The logic of this work is, therefore, as follows. We first present the factorized scalar and read off its zeros, poles, and singular factors. We then derive the same expression from the Brown--York membrane stress tensor on a stretched horizon, and clarify precisely what is meant by analytic continuation away from that surface. Next, we rewrite the scalar in a form adapted to the stationary and axial Killing fields. Finally, we briefly discuss two limitations of the construction: its relation to existing quasilocal-stress formulas and the status of the effective-fluid analogy. The derivational details are collected in the Appendix.

\section{A factorized scalar for distinguished Kerr--Newman surfaces}
Consider the Kerr–Newman metric in Boyer–Lindquist coordinates $(t,r,\theta,\phi)$,
\begin{flalign}
\label{eq:KN_metric}
    ds^2=&-\left(1-\frac{2mr-q^2}{\Sigma}\right)dt^2+\frac{\Sigma}{\Delta}dr^2 \\
    &-2a\,\frac{2mr-q^2}{\Sigma}\sin^2\theta \,dt\, d\phi+\Sigma\, d\theta^2 \notag\\
    &+\left(r^2+a^2+\frac{2mr-q^2}{\Sigma}a^2\sin^2\theta\right)\sin^2\theta\, d\phi^2,\notag
\label{eq:KN_metric_secondline}
\end{flalign}
where $(m,a,q)$ are the mass, rotation, and electric charge, respectively, and
\begin{equation}
\Sigma:=r^2+a^2\cos^2\theta,
\qquad
\Delta:=r^2+a^2-2mr+q^2.
\label{eq:SigmaDelta}
\end{equation}
The quantity that will play the central role in this work is
\begin{equation}
\small
P_{\rm KN}(r,\theta)
=
\frac{
 m\,(r-r_{\rm h}^+)(r-r_{\rm h}^-)(r-\rho^+)(r-\rho^-)
}{
 8\pi\,(r-r_{\rm ergo}^+)(r-r_{\rm ergo}^-)
 \bigl(r^2+|\rho^+\rho^-|\bigr)^2
},
\label{eq:PKN_letter}
\end{equation}
where
\begin{equation}
r_{\rm h}^{\pm}:=m\pm\sqrt{m^2-a^2-q^2},
\label{eq:horizons_letter}
\end{equation}
\begin{equation}
r_{\rm ergo}^{\pm}:=m\pm\sqrt{m^2-a^2\cos^2\theta-q^2},
\label{eq:ergosurfaces_letter}
\end{equation}
and
\begin{equation}
\rho^{\pm}:=
\frac{q^2\pm\sqrt{q^4+4a^2m^2\cos^2\theta}}{2m}.
\label{eq:rho_pm_letter}
\end{equation}
The geometric content of \eqref{eq:PKN_letter} can be read off directly from its factorized form. The zeros at
\begin{equation}
r=r_{\rm h}^{\pm}
\label{eq:P_zero_horizons_letter}
\end{equation}
identify the outer event horizon and the inner Cauchy horizon. The simple poles at
\begin{equation}
r=r_{\rm ergo}^{\pm}
\label{eq:P_poles_ergo_letter}
\end{equation}
identify the outer and inner stationary-limit surfaces. The singular behavior associated with the ring singularity is encoded in the final denominator factor, since
\begin{equation}
\rho^+\rho^-=-a^2\cos^2\theta,
\qquad
r^2+|\rho^+\rho^-|=r^2+a^2\cos^2\theta=\Sigma,
\label{eq:Sigma_identity_letter}
\end{equation}
so that
\begin{equation}
P_{\rm KN}(r,\theta)\sim \Sigma^{-2}
\qquad
\text{as}
\qquad
\Sigma\to0.
\label{eq:ring_behavior_letter}
\end{equation}
Finally, as the Kerr–Newman metric is asymptotically flat, one has
\begin{equation}
P_{\rm KN}(r,\theta)\to0
\qquad
\text{as}
\qquad
r\to\infty.
\label{eq:asymptotics_letter}
\end{equation}
Thus, the analytic structure of this scalar displays, in one expression, the horizon factors, the stationary-limit factors, the ring-singularity factor, and the large-distance decay.

This compact organization should be interpreted carefully. It is not intended to replace invariant horizon-detection methods, nor to claim that the horizons, ergosurfaces, and singularity are newly discovered. Horizons and ergosurfaces can be detected by suitably chosen curvature invariants or Cartan invariants, but those constructions are usually tailored to particular surfaces \cite{Karlhede,Abdel,Page,McNutt,Tavlayan}. For example, the scalar detecting the event horizon of the Kerr black hole in the construction of Abdelqader and Lake \cite{Abdel} is a non-trivial combination of quadratic and differential invariants. Equation \eqref{eq:PKN_letter} has a different purpose: it provides a membrane-pressure-based scalar whose simple factorized form collects several standard Kerr--Newman loci in one place. It would be interesting to ask whether analogous pressure-derived scalar representatives exist for other black-hole families admitting a comparable stretched-horizon decomposition. We do not pursue this question here.

\section{Origin from the analytically continued membrane pressure}

The quantity \eqref{eq:PKN_letter} is not introduced as an arbitrary function. Its origin lies in the membrane paradigm, in which the null event horizon is replaced by a fictitious timelike stretched horizon endowed with fluid properties \cite{Damour,membrane_THORNE,TheMembraneModelofBlackHolesandApplications}. The quantity used below is a radial pressure representative associated with the Brown--York-type membrane stress tensor \cite{BrownYork,BrownLauYork,Wilczek}. For rotating black holes of Kerr type, one may use a Parikh--Wilczek/ADM-type decomposition \cite{Wilczek,Ulus,Agca:2024tqn}, for which the relevant seed functions are
\begin{equation}
F_t=\sqrt{1-\frac{2mr-q^2}{\Sigma}},
\qquad
F_r=\sqrt{\frac{\Sigma}{\Delta}}.
\label{eq:FtFr_letter}
\end{equation}
On the stretched horizon, the associated radial pressure representative takes the simple form
\begin{equation}
P_{\rm KN}=
\frac{1}{8\pi F_r^{\,2}}\,\partial_r\ln F_t.
\label{eq:P_membrane_letter}
\end{equation}
Our key step is mathematical rather than physical: after deriving \eqref{eq:P_membrane_letter} on a stretched horizon, we use the same algebraic expression as a scalar representative wherever the metric functions define an analytic continuation. This does not mean that the membrane paradigm is being physically extended into the black-hole interior, nor that there is a real membrane fluid inside the horizon. It only means that the scalar formula inherited from the stretched-horizon pressure is continued and then studied as a diagnostic function on the Kerr--Newman coordinate domain.

For the Kerr–Newman black hole, substituting \eqref{eq:FtFr_letter} into \eqref{eq:P_membrane_letter} yields
\begin{equation}
P_{\rm KN}
=
\frac{\Delta\Bigl[m\bigl(r^2-a^2\cos^2\theta\bigr)-rq^2\Bigr]}
{8\pi\Sigma^2\bigl(\Sigma-2mr+q^2\bigr)},
\label{eq:P_unfactorized_letter}
\end{equation}
which, factorizes exactly into \eqref{eq:PKN_letter}. Hence, the factorized scalar is the analytically continued membrane-pressure expression expressed in a form in which its zeros, poles, and singular factors are explicit. The Appendix explains how \eqref{eq:P_membrane_letter} arises from the Brown--York membrane stress tensor and why the same expression admits a natural covariant rewriting on the two-dimensional orbit space of the stationary and axial Killing fields.

It is useful to emphasize the relation to the quasilocal stress-tensor literature. Brown and York derived the quasilocal stress tensor on a timelike boundary and its decomposition into energy, momentum, and spatial stress components \cite{BrownYork}. Brown, Lau, and York later developed the action-and-energy interpretation of these boundary quantities in a form especially suited to gravitational thermodynamics \cite{BrownLauYork}. Our Eq.~\eqref{eq:P_membrane_letter} is not a new quasilocal stress tensor. It is the pressure component obtained from the standard Brown--York membrane stress tensor after the stretched-horizon split described in the Appendix. Equation~\eqref{eq:P_geometric_letter} below is then only a Killing-field rewriting of this pressure expression.

\section{Geometric rewriting in terms of Killing data}
The same expression also admits a geometrically transparent rewriting in terms of the stationary Killing vector $\xi^\mu=(\partial_t)^\mu$. Introduce the two commuting Killing vectors
\begin{equation}
\xi=\frac{\partial}{\partial t},
\qquad
\psi=\frac{\partial}{\partial \phi}.
\label{eq:KN_killing_vectors}
\end{equation}
Then, the metric \eqref{eq:KN_metric} may be written as
\begin{equation}
ds^2=
\frac{\Sigma}{\Delta}\,dr^2
+\Sigma\,d\theta^2
+\xi^2\,dt^2
+2(\xi\!\cdot\!\psi)\,dt\,d\phi
+\psi^2\,d\phi^2.
\label{eq:KN_metric_killing_compact}
\end{equation}
Since
\begin{equation}
F_t^2=-\xi\!\cdot\!\xi,
\qquad
F_r^2=g_{rr},
\label{eq:FtFr_geometric_letter}
\end{equation}
Eq.~\eqref{eq:P_membrane_letter} becomes
\begin{equation}
P_{\rm KN}=
\frac{1}{16\pi}\,g^{rr}\,\partial_r\ln(-\xi\!\cdot\!\xi).
\label{eq:P_geometric_letter}
\end{equation}
Thus, the continued membrane-pressure scalar is governed by the radial variation of the norm of the stationary Killing field. This form should be read as a rewriting of the Brown--York membrane pressure, not as an independent quasilocal construction.

This expression can be written in a more covariant form by using the two-dimensional orbit space orthogonal to the stationary and axial Killing orbits. Let
\begin{equation}
K_A^\mu=(\xi^\mu,\psi^\mu),
\quad
\mathcal G_{AB}:=K_A\!\cdot K_B,
\quad
\mathcal G^{AB}:=(\mathcal G^{-1})^{AB},
\label{eq:Gram_defs}
\end{equation}
and define the projector onto the orbit space by
\begin{equation}
\Pi^\mu{}_{\nu}
=
\delta^\mu{}_{\nu}
-
K_A^{\mu}\,\mathcal G^{AB}\,K_{B\nu}.
\label{eq:orbit_projector_short}
\end{equation}
For any scalar $f$ invariant under the Killing flows, define its orbit-space derivative by
\begin{equation}
D^{\mu}f:=\Pi^{\mu\nu}\nabla_{\nu}f.
\label{eq:orbit_D_defs}
\end{equation}
If $\rho(x)$ is a scalar whose level sets define the membrane foliation, then the pressure may be written as
\begin{equation}
P_{\rm KN}
=
\frac{1}{16\pi}\,D^{\mu}\rho\,\nabla_{\mu}\ln(-\xi\!\cdot\!\xi).
\label{eq:P_covariant_h_form}
\end{equation}
For the Boyer--Lindquist choice $\rho=r$, one recovers $D^{\mu}r\,\nabla_{\mu}=g^{rr}\partial_r$, and hence Eq.~\eqref{eq:P_geometric_letter}. This shows that the analytically continued membrane-pressure scalar is the transverse variation, within the orbit space, of the norm of the stationary Killing field. The stationary-limit surfaces appear as poles precisely because they are the loci at which $\xi^\mu$ becomes null. The Appendix gives the corresponding orbit-space construction in a slightly more expanded form.

\section{Meaning of the $\rho^\pm$ surfaces}
Unlike the horizons and stationary-limit surfaces, the radii $\rho^\pm$ are not causal boundaries. Their origin is instead tied to the radial behavior of the norm of the stationary Killing field $\xi^\mu$. Outside the ergoregion, where $\xi^\mu$ is timelike, its integral curves describe the static observers at fixed Boyer--Lindquist spatial coordinates. The following observer interpretation is restricted to that domain; inside the ergoregion no such static observer exists, and the continued scalar should be interpreted only algebraically. Their normalized four-velocity is
\begin{equation}
u^\mu:=\frac{\xi^\mu}{\sqrt{-\xi\!\cdot\!\xi}},
\label{eq:u_static_rho}
\end{equation}
and their four-acceleration satisfies
\begin{equation}
a_{\mu}:=u^\nu\nabla_\nu u_{\mu}
=\frac{1}{2}\nabla_{\mu}\ln(-\xi\!\cdot\!\xi).
\label{eq:a_static_rho}
\end{equation}
Thus, the support force required to keep an observer static is governed directly by the gradient of the Killing norm. For the Kerr–Newman metric, one has
\begin{equation}
-\xi\!\cdot\!\xi
=1-\frac{2mr-q^2}{\Sigma},
\qquad
\Sigma=r^2+a^2\cos^2\theta,
\label{eq:xi_norm_rho}
\end{equation}
and therefore
\begin{equation}
\partial_r(-\xi\!\cdot\!\xi)
=
\frac{2\bigl[m(r^2-a^2\cos^2\theta)-rq^2\bigr]}{\Sigma^2}.
\label{eq:dr_xi_norm_rho}
\end{equation}
Hence, the condition
\begin{equation}
\partial_r(\xi\!\cdot\!\xi)=0
\label{eq:radial_extremum_condition}
\end{equation}
is equivalent to
\begin{equation}
m(r^2-a^2\cos^2\theta)-rq^2=0,
\label{eq:rho_quadratic}
\end{equation}
whose roots are precisely
\begin{equation}
r=\rho^\pm
=\frac{q^2\pm\sqrt{q^4+4a^2m^2\cos^2\theta}}{2m}.
\label{eq:rho_pm_meaning}
\end{equation}
In the region where static observers exist, the surfaces $r=\rho^\pm(\theta)$ therefore mark the loci at which the radial derivative of the stationary redshift factor vanishes. Equivalently, they mark where the radial component of the support acceleration of the static congruence changes sign. In this restricted sense, $\rho^\pm$ define angle-dependent balance surfaces of the stationary geometry. They are not horizons, and they do not render the static observers geodesic in the full spacetime; rather, they identify where the radial part of the force needed to maintain staticity reverses direction. Outside the static-observer domain, the same factors are retained only through the analytic continuation of $P_{\rm KN}$.

This interpretation is also reflected directly in the continued scalar. The same quadratic factor that defines $\rho^\pm$ appears in the numerator of \eqref{eq:P_unfactorized_letter}, so $P_{\rm KN}$ vanishes at these surfaces. In the static exterior region, the sign change of $P_{\rm KN}$ across $r=\rho^\pm$ mirrors the sign change in the radial support force of the static observers. For $m>0$, one has $\rho^+\ge0$ and $\rho^-\le0$. Therefore, on the positive-$r$ branch, $\rho^+(\theta)$ is the nonnegative root, while $\rho^-$ belongs to the analytically continued negative-$r$ branch of the full Kerr--Newman expression. A direct physical static-observer interpretation is available only where $\xi^\mu$ is timelike.

\section{Secondary effective-fluid interpretation}
We now record a secondary algebraic observation. This observation requires a simple dictionary choice between the radial geometric variable and the specific-volume variable. The same continued scalar can be formally read as an effective equation of state, with the radial variable $r$ playing the role of a specific-volume variable,
\begin{equation}
v:=r,
\label{eq:v_equals_r_letter}
\end{equation}
 This is only a formal analogy: we do not claim that the Kerr--Newman geometry is literally modeled by an ordinary thermodynamic fluid, nor that the continued pressure defines a physical equation of state in the black-hole interior.

The only point is that the analytic structure of $P_{\rm KN}$ resembles that of a generalized multi-component van der Waals expression. In particular, the poles at the stationary-limit surfaces are analogous to excluded-volume loci, while the large-$r$ inverse-power terms resemble virial corrections. This analogy is already visible from the factorized form \eqref{eq:PKN_letter}, and becomes especially transparent in local expansions. For example, near the outer stationary-limit surface,
\begin{equation}
r=r_{\rm ergo}^+ + \varepsilon,
\qquad
\varepsilon\to0,
\label{eq:r_outer_ergo_eps_letter}
\end{equation}
the pressure admits the Laurent expansion
\begin{equation}
P_{\rm KN}(r,\theta)
=
\frac{A_{-1}(\theta)}{r-r_{\rm ergo}^+}
+
A_0(\theta)
+
{\cal O}(r-r_{\rm ergo}^+).
\label{eq:P_outer_ergo_Laurent_letter}
\end{equation}
This has the same local algebraic form as a van der Waals-type pressure,
\begin{equation}
P(v,\theta)
=
\frac{T^{\rm eff}_{+,\rm ergo}(\theta)}{v-b}
+
P^{\rm reg}_{+,\rm ergo}(\theta)
+\cdots,
\qquad
b=r_{\rm ergo}^+,
\label{eq:local_vdw_letter}
\end{equation}
with the pole location playing the role of an effective excluded-volume scale and the residue playing the role of an effective, geometry-dependent temperature parameter. This interpretation is not used in any of the geometric claims above.

\begin{figure}[!]
\centering
\subfigure{%
\label{fig1}%
\includegraphics[height=2in]{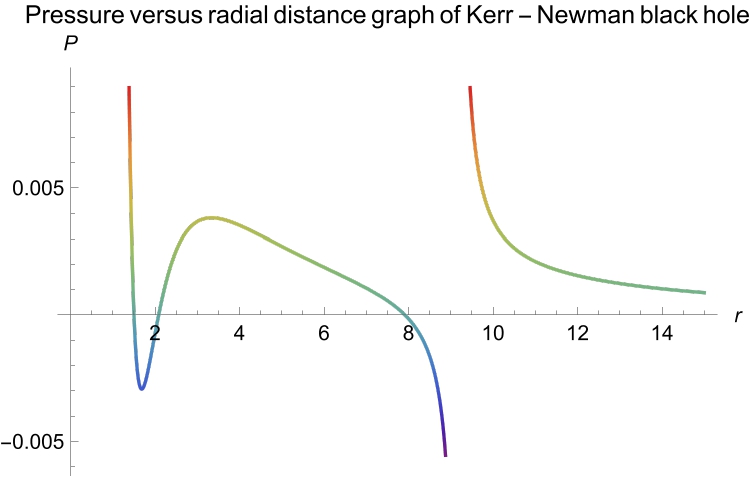}}%
\label{fig:third}%
\caption{Equatorial plot of the analytically continued pressure $P(r)$ for Kerr--Newman  black hole. Here we set $m=5$, $a=3$, $q=2.74$, and $\theta=\pi/2$. }
\label{figure1}
\end{figure}
\begin{figure}[!]
\centering
\subfigure{%
\label{fig2}%
\includegraphics[height=2in]{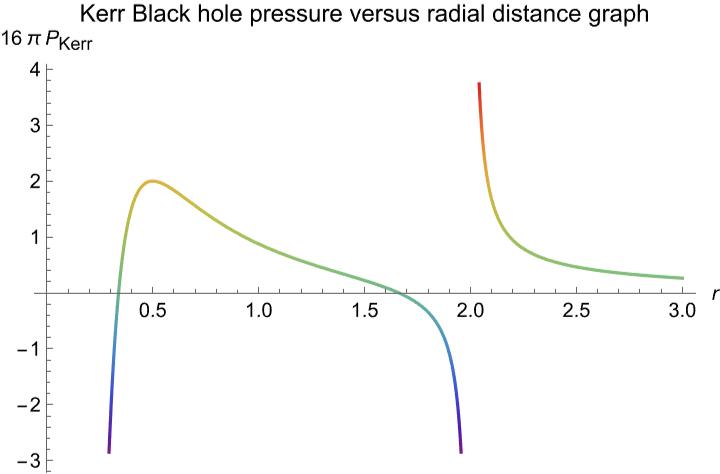}}%
\qquad
\subfigure{%
\label{fig:fourth}%
\includegraphics[height=2in]{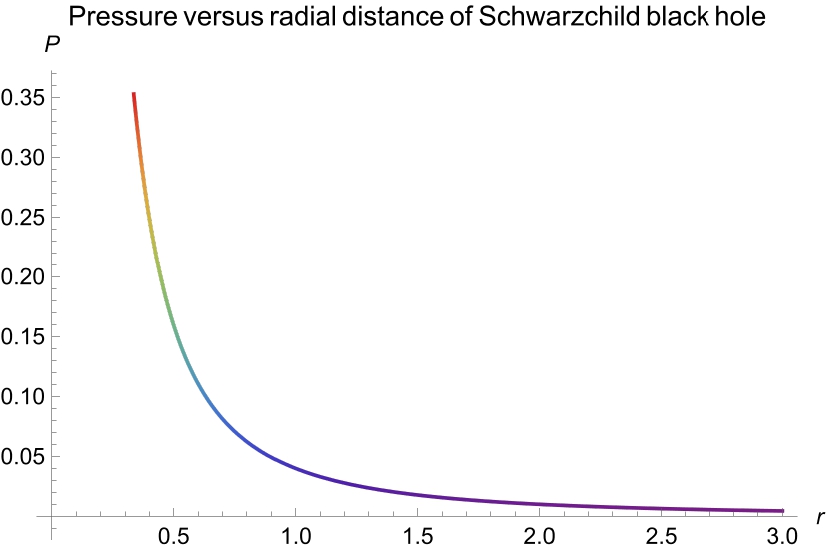}}%
\caption{Equatorial plots illustrating the Kerr and Schwarzschild limits of the pressure function. In the upper panel, we set $m=1$ and $a=0.75$ for the Kerr case. The zeros coincide with the Kerr horizons, while the pole marks the stationary-limit surface. In the lower panel, the $a=0$ Schwarzschild limit is shown: the stationary-limit surface merges with the horizon, and the plot retains only the horizon zero together with the asymptotic decay.}
\label{figure2}
\end{figure}

The geometric content of \eqref{eq:PKN_letter} is illustrated in Figs.~\ref{figure1} and \ref{figure2}. These plots are not used as evidence for the analytic claims, which already follow from the exact factorized formula, but they provide a useful visualization of how zeros, poles, and the large-$r$ falloff appear in representative slices and limits of the solution. Figure~\ref{figure1} displays the full Kerr--Newman behavior on the equatorial plane, while Fig.~\ref{figure2} shows the Kerr and Schwarzschild limits.

\section{Conclusion and discussion}

We have shown that the Kerr--Newman black hole admits a compact scalar representative, $P_{\rm KN}(r,\theta)$, obtained by analytically continuing the stretched-horizon membrane pressure function. In fully factorized form, this scalar has zeros at the outer and inner horizons, poles at the outer and inner stationary-limit surfaces, a higher-order singular factor associated with $\Sigma=r^2+a^2\cos^2\theta$, and decay at large $r$. The result should be understood as a compact reorganization of known Kerr--Newman surface data, not as a new invariant characterization of those surfaces.

We have also explained the construction's origin and limitations. The scalar arises from the Brown--York membrane stress tensor on a stretched horizon. The continuation away from the stretched horizon is an analytic continuation of the resulting scalar expression, not a physical assertion that a membrane fluid exists inside the horizon or inside the ergoregion. With this caveat, the expression is geometrically natural: it is governed by the transverse variation, in the orbit space, of the norm of the stationary Killing field.

We have also recorded a limited effective-fluid analogy. The continued scalar has a van der Waals-type algebraic structure near its pole loci, but this analogy is secondary and should not be confused with an ordinary thermodynamic equation of state for the black hole.

It is useful to compare this construction with other approaches to distinguished Kerr--Newman surfaces. Pugliese and Quevedo \cite{PuglieseQuevedo}, for example, studied the relations between black holes and naked singularities via horizon remnants, Killing throats, and bottlenecks. Their analysis is broader in parameter space, whereas the present work is narrower: it provides an explicit scalar inherited from the membrane pressure and studies how its factorized analytic structure aligns with standard surfaces in a given Kerr--Newman geometry. Natural extensions include expressing the numerator and denominator factors in terms of curvature or Cartan invariants, studying less symmetric stationary black holes, and clarifying whether the formal effective-fluid analogy can be made useful without obscuring the main geometric content.
\section*{Acknowledgments}
We thank Manus Visser for useful discussions.

\onecolumngrid
\section*{Appendix: Membrane-paradigm origin and geometric form of the pressure}
\label{Endnote:How to find the generic pressure formula}
This Appendix summarizes the derivation steps underlying the pressure formula used in the main text and makes explicit its relation to the Brown--York quasilocal stress tensor. Its purpose is twofold. First, it explains how the membrane-paradigm expression
\begin{equation}
P=\frac{1}{8\pi F_r^2}\,\partial_r\ln F_t
\end{equation}
arises for Kerr-type rotating black holes. Second, it shows how this coordinate-adapted formula may be rewritten in a covariant form on the two-dimensional orbit space used in the main text.

\subsection*{1. Geometric set-up}

Consider a timelike stretched horizon \(\mathcal M\) with an outward-pointing spacelike unit normal \(n^a\), and let the spacetime metric be decomposed as
\begin{align}
g_{ab} = h_{ab} + n_a n_b,
\end{align}
where \(h_{ab}\) is the induced metric on the stretched horizon. Inside \(\mathcal M\), one further separates the timelike direction \(u^a\) and the spatial \((d-2)\)-dimensional metric \(\gamma_{ab}\), so that
\begin{align}
h_{ab} = - u_a u_b + \gamma_{ab},
\qquad
u^a u_a=-1,
\qquad
u^a \gamma_{ab}=0.
\end{align}
This is the \((d-2)+1+1\) split adapted to the stretched horizon.

Upon variation, the action principle should be satisfied on the boundary, which leads to the quasi-local Brown–York--type membrane stress tensor defined at a finite radius, a time-like boundary
\begin{align}
t_{ab} = \frac{1}{8\pi G}\left(K h_{ab} - K_{ab}\right),
\label{eq:endnote_BY_tensor}
\end{align}
where $K_{ab}=h_a{}^c \nabla_c n_b$ is the extrinsic curvature of the timelike stretched horizon, and $K=h^{ab}K_{ab}$
is its trace. This is the standard Brown--York boundary stress tensor, specialized here to the stretched-horizon timelike boundary. The pressure extracted below is therefore a component of the usual quasilocal stress tensor, not a new boundary stress tensor.

\subsection*{2. Decomposition of the stretched-horizon extrinsic curvature}
The extrinsic curvature could be decomposed further into the spatial section $\gamma_{ab}$, mixed terms, and purely time-like terms. Following the decomposition $h_{ab}=-u_a u_b+\gamma_{ab}$, one arrives at the result.
\begin{align}
K_{ab}
=
k_{ab}
-u_a \Omega_b
-u_b \Omega_a
-g_{\mathcal H}\,u_a u_b,
\label{eq:endnote_K_decomp}
\end{align}
here, $k_{ab}:=\gamma_a{}^c \gamma_b{}^d \nabla_c n_d$ is the extrinsic curvature of the spatial \((d-2)\)-surface inside the stretched horizon, $\Omega_a:=\gamma_a{}^c u^d \nabla_c n_d$ is the momentum one-form on the membrane, and $g_{\mathcal H}:=n_a a^a,$ where $
a^a:=u^b \nabla_b u^a$ is the normal component of the acceleration of the fiducial observers. Taking the trace with \(h^{ab}\), one finds
\begin{equation}
K=k+g_{\mathcal H}.
\label{eq:endnote_K_trace}
\end{equation}
\subsection*{3. Membrane stress tensor in decomposed form}
Substituting \eqref{eq:endnote_K_decomp} and \eqref{eq:endnote_K_trace} into
\eqref{eq:endnote_BY_tensor}, one finds
\begin{equation}
t_{ab}=
\frac{1}{8\pi G}
\left[
(k+g_{\mathcal H})\gamma_{ab}
-k_{ab}
-k\,u_a u_b
+u_a\Omega_b
+u_b\Omega_a
\right].
\label{eq:endnote_t_decomp}
\end{equation}
Projecting this tensor along \(u^a\) and \(\gamma_{ab}\) gives the membrane energy density,
momentum density, and stress two-form:
\begin{flalign}
\rho &:= u^a u^b t_{ab} = -\frac{k}{8\pi G},
\quad\quad\quad\quad\quad\quad\quad\quad\quad\quad\pi_A := - \gamma_A{}^a u^b t_{ab} = \frac{\Omega_A}{8\pi G},
\label{eq:endnote_pi_generic}
\\
t_{AB}^{\text{(spatial)}} &:= \gamma_A{}^a \gamma_B{}^b t_{ab}
=
\frac{1}{8\pi G}\left[(k+g_{\mathcal H})\gamma_{AB}-k_{AB}\right].
\label{eq:endnote_spatialstress_generic}
\end{flalign}
We will be specifically interested in the spatial stress two-form on the $(d-2)$-dimensional spacelike cross-section of the stretched horizon, which will lead us to the pressure function given in the main text. 
\subsection*{4. Trace--traceless split and viscous-fluid identification}

Now decompose the spatial extrinsic curvature \(k_{AB}\) into trace and traceless parts:
\begin{align}
k_{AB}
=
\sigma_{AB}
+\frac{1}{d-2}\,\Theta\,\gamma_{AB},
\qquad
\gamma^{AB}\sigma_{AB}=0,
\label{eq:endnote_k_split}
\end{align}
where \(\Theta:=\gamma^{AB}k_{AB}\) is the expansion and \(\sigma_{AB}\) is the shear tensor of the stretched horizon.

Substituting \eqref{eq:endnote_k_split} into \eqref{eq:endnote_spatialstress_generic} gives
\begin{equation}
t_{AB}^{\text{(spatial)}}
=
\frac{1}{8\pi G}
\left[
-\sigma_{AB}
+
\left(
g_{\mathcal H}
+\frac{d-3}{d-2}\Theta
\right)\gamma_{AB}
\right].
\label{eq:endnote_spatialstress_split}
\end{equation}
To compare with a viscous fluid, write the spatial stress in standard form
\begin{align}
t_{AB}^{\text{(spatial)}}
=
p\,\gamma_{AB}
-2\eta\,\sigma_{AB}
-\zeta\,\Theta\,\gamma_{AB}.
\label{eq:endnote_viscous_form}
\end{align}
Matching \eqref{eq:endnote_spatialstress_split} with \eqref{eq:endnote_viscous_form}, one identifies
\begin{align}
p=\frac{g_{\mathcal H}}{8\pi G},
\qquad
\eta=\frac{1}{16\pi G},
\qquad
\zeta=-\frac{d-3}{8\pi G(d-2)}.
\label{eq:endnote_transport_generic}
\end{align}
In \(d=4\), this gives the standard membrane values
\begin{align}
\eta=\frac{1}{16\pi G},
\qquad
\zeta=-\frac{1}{16\pi G}.
\label{eq:endnote_transport_d4}
\end{align}

\subsection*{Static-observer acceleration and membrane pressure for the Kerr-type metric}

We now specialize the generic membrane-pressure formula to the Kerr-type metric written in the
Boyer–Lindquist type form.
\begin{align}
ds^{2}
&=
-f\,dt^{2}
+\frac{\Sigma}{\Delta}\,dr^{2}
+2a(f-1)\sin^{2}\theta\,dt\,d\phi
+\Sigma\,d\theta^{2}
+\Bigl[r^{2}+a^{2}-(1-f)a^{2}\sin^{2}\theta\Bigr]\sin^{2}\theta\,d\phi^{2},
\label{eq:Kerrtype_metric_appendix}
\end{align}
with
\begin{align}
\Sigma=r^{2}+a^{2}\cos^{2}\theta,
\qquad
f=1-\frac{2M(r)\,r}{\Sigma},
\qquad
\Delta=f\Sigma+a^{2}\sin^{2}\theta.
\label{eq:Kerrtype_functions_appendix}
\end{align}
For the Kerr--Newman spacetime, the mass function has the form $M(r)=m-\frac{q^{2}}{2r}
$, while the discriminant is  $\Delta=r^{2}+a^{2}-2mr+q^{2}$. Following the membrane-paradigm decomposition, introduce the functions
\begin{flalign}
F_t^{2}&=f,
\qquad\quad\quad\qquad\qquad\quad\,\,\,\,\,\,\,
F_r^{2}=\frac{\Sigma}{\Delta},\\
\omega&=-a(f-1)\sin^{2}\theta\,F_t^{-1},
\qquad
F_{\phi}^{2}
=
\Bigl[r^{2}+a^{2}+(1-f)a^{2}\sin^{2}\theta\Bigr]\sin^{2}\theta.
\label{eq:omegaFphi_appendix}
\end{flalign}
Then the metric may be rewritten as
\begin{align}
ds^{2}
=
-\bigl(F_t\,dt+\omega\,d\phi\bigr)^{2}
+F_r^{2}\,dr^{2}
+\Sigma\,d\theta^{2}
+\bigl(F_{\phi}^{2}+\omega^{2}\bigr)d\phi^{2}.
\label{eq:metric_rewritten_appendix}
\end{align}
The stretched-horizon membrane data are derived from the one-forms
\begin{align}
u_a dx^a = F_t\,dt+\omega\,d\phi,
\qquad
n_a dx^a = F_r\,dr,
\label{eq:u_n_oneforms_appendix}
\end{align}
On the stretched horizon, in the domain where the fiducial observers and the seed functions are well defined, we use the radial pressure representative

\begin{equation}
P_{\rm KN}
:=
\frac{1}{8\pi G}\,a^r
=
\frac{1}{8\pi G\,F_r^2}\,\partial_r\ln F_t .
\label{eq:endnote_pressure_FtFr}
\end{equation}
This is the scalar whose analytic continuation is studied in the main text.  It should not be confused with a physical membrane fluid continued into the black-hole interior.  The representative \eqref{eq:endnote_pressure_FtFr} can also be written in terms of the stationary Killing field as
\begin{equation}
P_{\rm KN}
=
\frac{1}{16\pi}g^{rr}\partial_r\ln(-\xi^2),
\end{equation}
where $\xi^\mu=(\partial_t)^\mu$ is the stationary Killing field. This is precisely the geometric form stated in Eq.~\eqref{eq:P_geometric_letter} of the main text.

\subsection*{4. Orbit-space form of the geometric pressure}
\label{Endnote: Orbit space}
A stationary axisymmetric spacetime admits two commuting Killing vector fields,
\begin{equation}
\xi^\mu=\left(\frac{\partial}{\partial t}\right)^\mu,
\qquad
\psi^\mu=\left(\frac{\partial}{\partial \phi}\right)^\mu.
\end{equation}
At each point where they are linearly independent, they span a two-dimensional Killing orbit. For circular spacetimes such as Kerr–Newman, the orthogonal complement of these orbits defines a two-dimensional orbit space
\begin{equation}
Q=M/\{\xi,\psi\},
\end{equation}
which, in Boyer–Lindquist coordinates, is coordinatized by
\begin{equation}
x^i=(r,\theta),
\qquad
y^A=(t,\phi).
\end{equation}
Define
\begin{equation}
K_A^\mu=(\xi^\mu,\psi^\mu),
\qquad
\mathcal G_{AB}:=K_A\!\cdot K_B,
\qquad
\mathcal G^{AB}:=(\mathcal G^{-1})^{AB}.
\end{equation}
The projector onto the orbit space is then
\begin{equation}
\Pi^\mu{}_{\nu}
=
\delta^\mu{}_{\nu}-K_A^\mu\,\mathcal G^{AB}\,K_{B\nu},
\end{equation}
and the induced metric on the orbit space is
\begin{equation}
q_{\mu\nu}
=
g_{\mu\nu}-K_{A\mu}\,\mathcal G^{AB}\,K_{B\nu}.
\end{equation}
In adapted coordinates, one may write
\begin{equation}
ds^2=\mathcal G_{AB}(x)\,dy^A dy^B + q_{ij}(x)\,dx^i dx^j,
\end{equation}
and for Kerr--Newman, one has
\begin{equation}
q_{ij}dx^i dx^j=\frac{\Sigma}{\Delta}\,dr^2+\Sigma\,d\theta^2.
\end{equation}
For any scalar $f$ invariant under the Killing flows, define the orbit-space derivative
\begin{equation}
D_\mu f:=\Pi_\mu{}^\nu\nabla_\nu f,
\qquad
D^\mu f:=\Pi^{\mu\nu}\nabla_\nu f.
\end{equation}
Let $\rho(x)$ be a scalar on the orbit space whose level sets define the membrane foliation. Then the natural covariant form of the continued membrane pressure is
\begin{equation}
P=\frac{1}{16\pi}\,D^\mu\rho\,\nabla_\mu\ln(-\xi^2).
\label{eq:appendix_orbit_pressure}
\end{equation}
For the Boyer–Lindquist choice
\begin{equation}
\rho=r,
\end{equation}
one finds
\begin{equation}
D^\mu r\,\nabla_\mu=g^{rr}\partial_r,
\end{equation}
and therefore
\begin{equation}
P=\frac{1}{16\pi}\,g^{rr}\partial_r\ln(-\xi^2),
\end{equation}
which is Eq.~\eqref{eq:P_geometric_letter} of the main text.
Equation \eqref{eq:P_covariant_h_form} of the main text and Eq.~\eqref{eq:appendix_orbit_pressure} of the Appendix therefore have a simple geometric meaning: the analytically continued membrane pressure is the transverse variation, within the orbit space, of the norm of the stationary Killing field. This also explains why the stationary-limit surfaces appear as poles, since they are precisely the loci at which
\begin{equation}
\xi^2=0.
\end{equation}
Thus, the scalar used in this article is not an arbitrary function; it is the analytic continuation of the stretched-horizon membrane pressure, written in a form adapted to the geometry of the stationary and axial Killing orbits. Its physical membrane interpretation, however, remains tied to the stretched horizon itself.

\twocolumngrid


\begin{thebibliography}{99}

\bibitem{Finkelstein}
D.~Finkelstein,
Past-Future Asymmetry of the Gravitational Field of a Point Particle,
Phys. Rev. \textbf{110}, 965-967 (1958).

\bibitem{Penrose}
R.~Penrose,
``Gravitational collapse: The role of general relativity,''
Riv. Nuovo Cim. \textbf{1}, 252-276 (1969).

\bibitem{Kerr}
R.~P.~Kerr,
Gravitational field of a spinning mass as an example of algebraically special metrics,
Phys. Rev. Lett. \textbf{11}, 237 (1963).

\bibitem{Carter}
B.~Carter,
``Global structure of the Kerr family of gravitational fields,''
Phys. Rev. \textbf{174}, 1559-1571 (1968).

\bibitem{Penrose2}
R.~Penrose,
``Gravitational collapse and space-time singularities,''
Phys. Rev. Lett. \textbf{14}, 57-59 (1965).

\bibitem{Karlhede}
A.~Karlhede, U.~Lindstrom and J.~E.~Aman,
A note on a local effect at the Schwarzschild sphere,
Gen. Rel. Grav. \textbf{14}, 569 (1982).

\bibitem{Abdel}
M.~Abdelqader and K.~Lake,
Invariant characterization of the Kerr spacetime: Locating the horizon and measuring the mass and spin of rotating black holes using curvature invariants,
Phys. Rev. D \textbf{91}, no.8, 084017 (2015).

\bibitem{Page}
D.~N.~Page and A.~A.~Shoom,
Local Invariants Vanishing on Stationary Horizons: A Diagnostic for Locating Black Holes,
Phys. Rev. Lett. \textbf{114}, no.14, 141102 (2015).

\bibitem{McNutt}
D.~D.~McNutt, M.~A.~H.~MacCallum, D.~Gregoris, A.~Forget, A.~A.~Coley, P.~C.~Chavy-Waddy and D.~Brooks,
Cartan Invariants and Event Horizon Detection, Extended Version,
Gen. Rel. Grav. \textbf{50}, no.4, 37 (2018).
[erratum: Gen. Rel. Grav. \textbf{52}, no.1, 6 (2020)].

\bibitem{Tavlayan}
A.~Tavlayan and B.~Tekin,
Event horizon detecting invariants,
Phys. Rev. D \textbf{101}, no.8, 084034 (2020).

\bibitem{PuglieseQuevedo}
D.~Pugliese and H.~Quevedo,
Disclosing connections between black holes and naked singularities: horizon remnants, Killing throats and bottlenecks,
Eur. Phys. J. C \textbf{79}, 209 (2019).

\bibitem{Damour}
T.~Damour,
Black Hole Eddy Currents,
Phys. Rev. D \textbf{18}, 3598-3604 (1978).

\bibitem{membrane_THORNE}
K.~S.~Thorne, R.~H.~Price, and D.~A.~Macdonald,
\textit{Black Holes: The Membrane Paradigm}
(Yale University Press, London, 1986).

\bibitem{TheMembraneModelofBlackHolesandApplications}
N.~Straumann,
The membrane model of black holes and applications,
Lect. Notes Phys. \textbf{514}, 111 (1998).

\bibitem{BrownYork}
J.~D.~Brown and J.~W.~York, Jr.,
Quasilocal energy and conserved charges derived from the gravitational action,
Phys. Rev. D \textbf{47}, 1407 (1993).

\bibitem{BrownLauYork}
J.~D.~Brown, S.~R.~Lau and J.~W.~York, Jr.,
Action and energy of the gravitational field,
Annals Phys. \textbf{297}, 175 (2002).

\bibitem{Ulus}
C.~U.~Agca and B.~Tekin,
Membrane paradigm approach to the Johannsen--Psaltis black hole,
Phys. Rev. D \textbf{109}, no.10, 10 (2024).

\bibitem{Agca:2024tqn}
C.~U.~Agca and B.~Tekin,
Fluid-membrane descriptions of various black holes,
Eur. Phys. J. C \textbf{85}, no.8, 890 (2025).

\bibitem{Psaltis}
T.~Johannsen and D.~Psaltis,
A Metric for Rapidly Spinning Black Holes Suitable for Strong-Field Tests of the No-Hair Theorem,
Phys. Rev. D \textbf{83}, 124015 (2011).

\bibitem{Newman}
E.~T.~Newman, E.~Couch, K.~Chinnapared, A.~Exton, A.~Prakash and R.~Torrence,
Metric of a Rotating, Charged Mass,
J. Math. Phys. \textbf{6}, 918-919 (1965).

\bibitem{Wilczek}
M.~Parikh and F.~Wilczek,
An action for black hole membranes,
Phys. Rev. D \textbf{58}, 064011 (1998).

\end{thebibliography}
\end{document}